\documentclass{article}
\usepackage{spconf,amsmath,graphicx}
\usepackage{multirow}
\usepackage{subcaption}
\usepackage{amsfonts}
\usepackage{algorithmic}
\usepackage{algorithm}

\title{Text-to-speech synthesis based on latent variable conversion using diffusion probabilistic model and variational autoencoder}
\name{Yusuke Yasuda$^\dagger$ and Tomoki Toda$^\dagger$\thanks{This work was partly supported by JST CREST Grant Number JPMJCR19A3 and JSPS KAKENHI Grant Number JP21H05054.}}
\address{$^\dagger$Nagoya university}
\begin{document}
%
\maketitle
\vspace{-2mm}
\begin{abstract}
Text-to-speech synthesis (TTS) is a task to convert texts into speech. Two of the factors that have been driving TTS are the advancements of probabilistic models and latent representation learning. We propose a TTS method based on latent variable conversion using a diffusion probabilistic model and the variational autoencoder (VAE). In our TTS method, we use a waveform model based on VAE, a diffusion model that predicts the distribution of latent variables in the waveform model from texts, and an alignment model that learns alignments between the text and speech latent sequences. Our method integrates diffusion with VAE by modeling both mean and variance parameters with diffusion, where the target distribution is determined by approximation from VAE. This latent variable conversion framework potentially enables us to flexibly incorporate various latent feature extractors. Our experiments show that our method is robust to linguistic labels with poor orthography and alignment errors.
\end{abstract}
%
\section{Introduction}
\label{sec:intro}
Text-to-speech synthesis (TTS) is a task to convert texts into speech. Recent TTS methods achieved high naturalness in synthetic speech that is comparable to human speech \cite{DBLP:conf/icassp/ShenPWSJYCZWRSA18_short, DBLP:journals/corr/abs-1809-08895, DBLP:conf/icml/KimKS21}. One of the reasons that have been driving the performance of TTS is the advancement of probabilistic models such as autoregressive \cite{oord2016wavenet}, generative flow \cite{rezende2015variational}, and generative adversarial network \cite{goodfellow2014generative} models. Recently, the diffusion probabilistic model (DPM) \cite{DBLP:conf/nips/HoJA20} has been developed as another advanced probabilistic model, and it has been intensively studied in speech domains such as neural vocoders \cite{DBLP:conf/iclr/ChenZZWNC21}, and TTS \cite{jeong21_interspeech, chen21p_interspeech}. DPM is a particularly interesting approach for speech generation in terms of conditioning, because speech generation is about conditional distribution modeling. The iterative inference process of diffusion enables conditioning by various methods such as adaptive prior \cite{DBLP:conf/iclr/LeeKS0LMQ0YL22}, classifier guidance \cite{DBLP:conf/icml/KimKY22}, and iterative latent variable refinement \cite{levkovitch22_interspeech}. 

Another driving factor of the TTS advancement is latent representation learning. The use of the variational autoencoder (VAE) \cite{DBLP:journals/corr/KingmaW13} is a popular representation learning method for speech style modeling \cite{DBLP:conf/interspeech/KlimkovRRD19, DBLP:conf/iclr/HsuZWZWWCJCSNP19}. Designing a discrete latent space in VAE enables the capture of phonetic latent information in speech \cite{DBLP:conf/nips/OordVK17}, and a categorical latent space enables the use of a semisupervised approach to capture the desired latent information such as speech emotion \cite{DBLP:conf/iclr/HabibMSBSSKB20}.
Self-supervised learning (SSL) based on a pretrained representation is also applied to speech synthesis to utilize large-scale unpaired texts or speeches \cite{jia21_interspeech, DBLP:conf/interspeech/SiuzdakDRJ22}. Supervised fine-tuning in SSL enables the adoption of pretrained features to specific domains, and it is used to capture accent features for TTS \cite{9829304}.

TTS is expected to be improved by incorporating a diffusion probabilistic model and latent representation learning. However, diffusion models typically use a fixed variance, which results in a suboptimal likelihood \cite{DBLP:conf/icml/NicholD21}, and the application of diffusion to latent variable modeling, where the variance of latent variables is approximated by a model, is as yet not known. In addition, there is room to investigate alignment methods in conditional distribution modeling with diffusion, where alignments between the condition and the target are not known in TTS.

In this research, we propose a TTS method based on latent variable conversion. Inspired by a recent TTS method based on VAE \cite{DBLP:journals/corr/KingmaW13} incorporating a waveform model \cite{DBLP:conf/icml/KimKS21}, our method convert texts into waveforms via a latent space. We use diffusion to model the distribution of a latent representation in the VAE-based waveform model to convert texts into waveforms. To model a latent variable with an approximated distribution, we formulate the diffusion-based acoustic model capable of modeling both mean and variance parameters. We also propose an alignment model to align latent representations of texts and acoustic features. Our method enables us to use various latent representation potentially including SSL-based methods, but we limit our focus on the normal VAE in this paper. Our contributions are as follows: (1) we propose a TTS method that models the distribution of latent variables with diffusion taking both mean and variance parameters into consideration;
(2) Our experimental results imply that the modeling data in diffusion provides a more stable speech quality than modeling noise with less training time; (3) Our experimental results show that our method is robust to linguistic labels with poor orthography and alignment errors.

\vspace{-4mm}
\section{Diffusion probabilistic model}
\label{sec:diffusion}
\vspace{-3mm}
DPM \cite{DBLP:conf/nips/HoJA20} introduces $T$ latent variables $x_1\dots x_T$ to model the data distribution $p(x_0)$ and decomposes its variational lower bound $L$ into three terms.
\vspace{-2mm}
\small
\begin{align}
    L_T &= \mathrm{KL}[q(x_T|x_0)\|p(x_T)] \label{eq:diff_T}\\
    L_{1:T-1} &= \sum_{t=2}^{T}L_{t-1},\label{eq:diff_t}\nonumber\\
    (L_{t-1} &= \mathrm{KL}[q(x_{t-1}|x_t,x_0)\|p(x_{t-1}|x_t)])\\
    L_0 &= \log p(x_0|x_1) \label{eq:diff_0}
\vspace{-2mm}
\end{align}
\normalsize
Here, $0 \le t \le T$ is diffusion time and KL means Kullback–Leibler divergence. Eq.~(\ref{eq:diff_T}) is a loss term of the noise distribution $p(x_T)$ at the last diffusion time, Eq.~(\ref{eq:diff_t}) is the sum of loss terms of each diffusion time, and Eq.~(\ref{eq:diff_0}) is a loss term of the data distribution $p(x_0|x_1)$ at the first diffusion time. Given the noise distribution $p(x_T)$, diffusion model distribution $p(x_{t-1}|x_t)$, and data distribution $p(x_0|x_1)$, data $x_0$ can be generated by reversing diffusion time starting from noise $x_T$ via latent variables $x_{T-1}\dots x_1$ as the inference of DPM. The training of DPM involves optimizing objective functions of Eqs.~(\ref{eq:diff_T}), (\ref{eq:diff_t}), and (\ref{eq:diff_0}) to learn probability distributions.

In practice, the training of DPM can be simplified. If the noise distribution $p(x_T)$ is assumed to be a standard Gaussian distribution, Eq.~(\ref{eq:diff_T}) becomes constant and is ignored. Furthermore, modeling the data distribution $p(x_0|x_1)$ can be omitted by using mean parameter as a sample instead of random sampling. With these simplifications, the loss term to be optimized is only Eq.~(\ref{eq:diff_t}) to learn the distribution $p_\theta(x_{t-1}|x_t)$. In addition, we can partially optimize Eq.~(\ref{eq:diff_t}) by using a single term $L_{t-1}$ in summation when the diffusion time $t$ is randomly sampled.

During the inference of DPM, data $x_0$ can be obtained via sampling $x_{T}\dots x_1$ iteratively from the learned distribution $p_\theta(x_{t-1}|x_t)$ staring from noise $x_T$ as the initial input.

The derivation of latent variables $x_t$ and the approximate posterior $q(x_{t-1}|x_t,x_0)$ in Eq.~(\ref{eq:diff_t}) starts with the definition of a forward diffusion kernel $q(x_t|x_{t-1})$. If we define the forward diffusion kernel as $q(x_t|x_{t-1};\beta_t) = \mathcal{N}(x_t|\sqrt{1-\beta_t}x_{t-1}, \beta_t I)$, we obtain a distribution of latent variables $x_t$ as $q(x_t|x_0) = \mathcal{N}(x_t|\sqrt{\bar{\alpha}_{t}}x_0, (1-\bar{\alpha}_{t}) I)$, where $\beta_t$ is a variance configured with an arbitrary value, and we use $\alpha_t = 1 - \beta_t, \bar{\alpha}_t = \prod_{s=1}^t\alpha_s$ for visibility. 
With this distribution, the parameters of the approximate posterior distribution $q(x_{t-1}|x_t,x_0)=\mathcal{N}(x_{t-1}|\bar{\mu}(x_t,x_0),\bar{\sigma}^2(t))$ becomes:
\small
\begin{align}
    \bar{\mu}(x_t,x_0) &= \frac{\sqrt{\bar{\alpha}_{t-1}}\beta_t}{1-\bar{\alpha}_t}x_0 + \frac{\sqrt{\alpha_t}(1 - \bar{\alpha}_{t-1})}{1-\bar{\alpha}_t}x_t \label{eq:mean-approx-poasterior}\\
    \bar{\sigma}^2(t) = \bar{\beta}_t &= \frac{1-\bar{\alpha}_{t-1}}{1-\bar{\alpha}_t}\beta_t\label{eq:variance-approx-poasterior}
\end{align}
\normalsize

There are two ways to implement the model distribution \\$p_\theta(x_{t-1}|x_t)$ in Eq.~(\ref{eq:diff_t}) with the model function $f_\theta(x_t, t)$ \cite{DBLP:journals/corr/abs-2207-06389}. Let $p_\theta(x_{t-1}|x_t)$ be the Gaussian distribution with the mean $\mu_\theta(x_t,t)$ and variance $\sigma_\theta^2(x_t,t)$ parameters. The implementation choices are whether the model function predicts data $x_0$ or noise $\epsilon \sim \mathcal{N}(0, I)$. 
When modeling data $x_0$ with a model function $x_0 = f_\theta(x_t, t)$, we can obtain $\mu_\theta(x_t,t)$ in the model distribution $p_\theta(x_{t-1}|x_t)$ on the basis of on Eq.~(\ref{eq:mean-approx-poasterior}) by setting $\mu_\theta(x_t,t) = \frac{\sqrt{\bar{\alpha}_{t-1}}\beta_t}{1-\bar{\alpha}_t}f_\theta(x_t, t) + \frac{\sqrt{\alpha_t}(1 - \bar{\alpha}_{t-1})}{1-\bar{\alpha}_t}x_t$. 
When modeling noise $\epsilon$ with a model function $\epsilon = f_\theta(x_t, t)$, we can use the relationship between $\epsilon$ and $x_0$ as $x_t = \sqrt{\bar{\alpha}_t}x_0 + \sqrt{1-\bar{\alpha}_t}\epsilon$ used in sampling $x_t$ from $q(x_{t-1}|x_0;\beta)$. 
We can therefore obtain the model distribution $p_\theta(x_{t-1}|x_t)$ with the mean parameter $\mu_\theta(x_t,t) = \frac{1}{\sqrt{\alpha_t}}x_t - \frac{\beta_t}{\sqrt{\alpha_t}\sqrt{1-\bar{\alpha}_t}}f_\theta(x_t,t)$ on the basis of Eq.~(\ref{eq:mean-approx-poasterior}). The noise prediction is equivalent to score matching \cite{DBLP:conf/nips/SongE19}.

\vspace{-4mm}
\section{Proposed method}
\vspace{-2mm}
\label{sec:method}
\begin{figure}[!t]
    \begin{center}
    \begin{subfigure}[t]{0.7\columnwidth}
    {\includegraphics[width=\columnwidth]{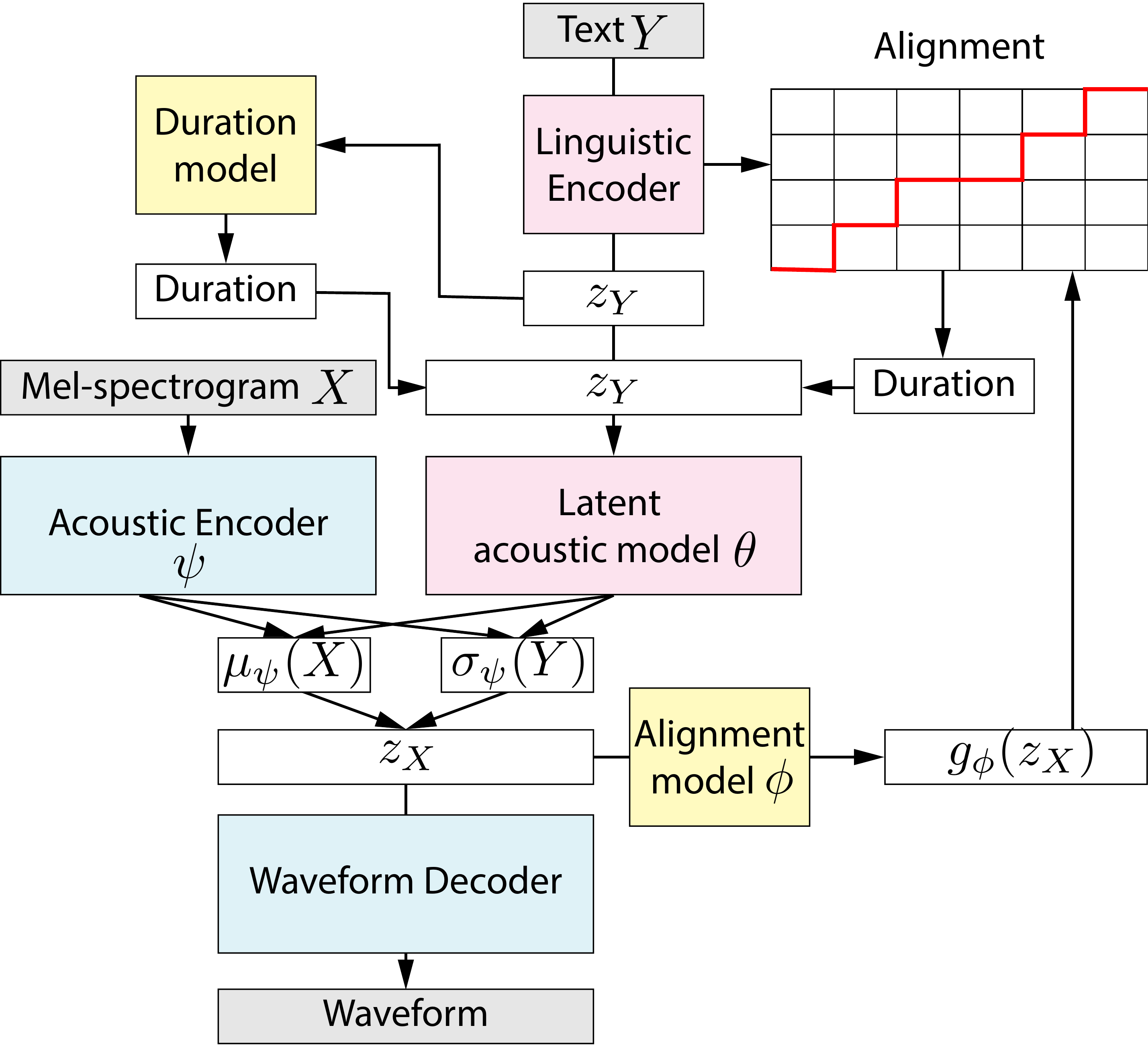}}
    \subcaption{Training}
    \end{subfigure}
    \begin{subfigure}[t]{0.5\columnwidth}
    {\includegraphics[width=\columnwidth]{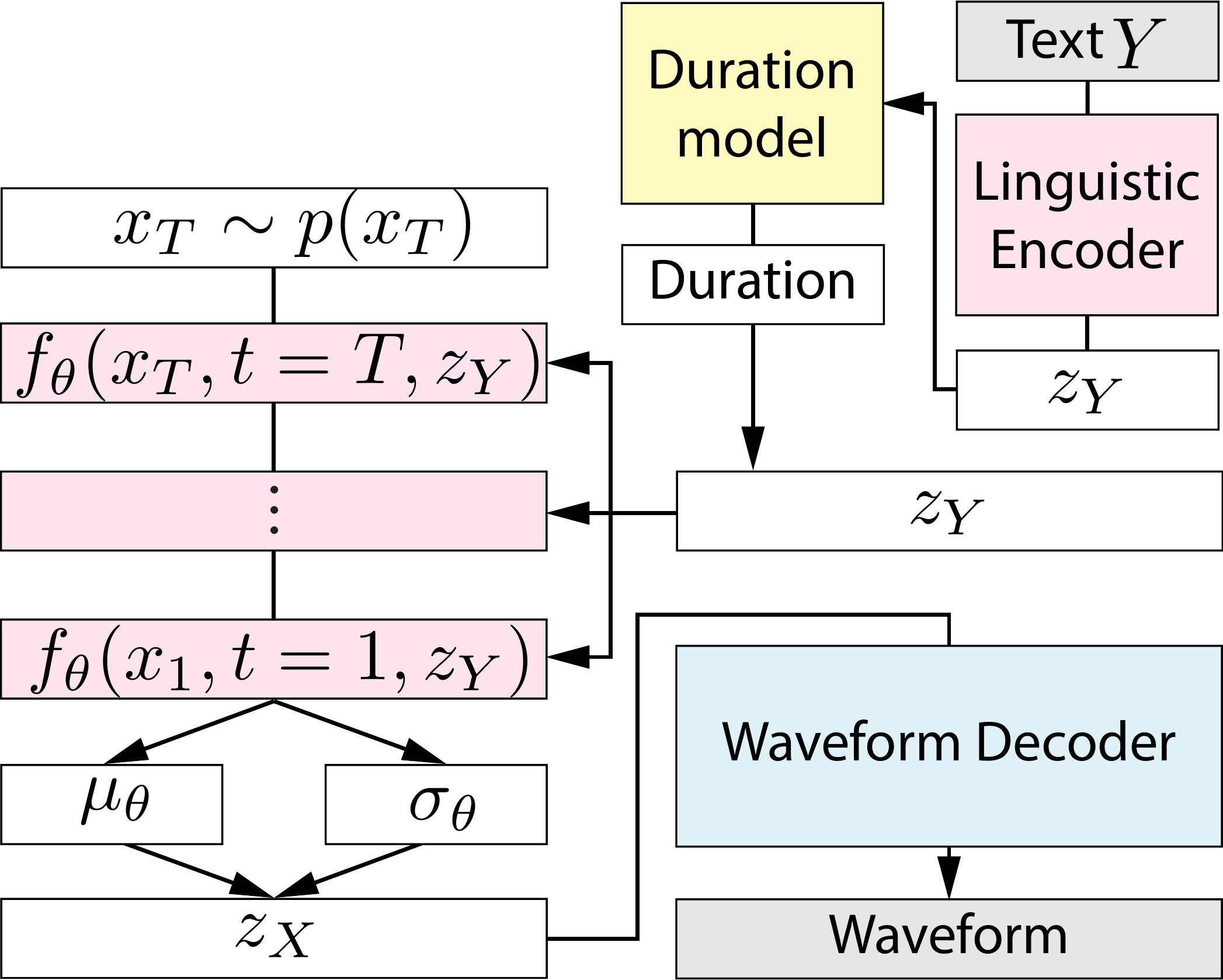}}
    \subcaption{Inference}
    \vspace{-2mm}
    \end{subfigure}
    \end{center}
    \vspace{-4mm}
    \caption{Proposed TTS method based on diffusion and VAE.}
    \label{fig:diffusion-tts}
    \vspace{-6mm}
\end{figure}
Figure~\ref{fig:diffusion-tts} shows our proposed TTS method based on DPM. Our proposed method consists of three parts: the (1) waveform, (2) latent acoustic, and (3) alignment models. 
\vspace{-4mm}
\subsection{Training and inference}
\vspace{-2mm}
\noindent 
\textbf{Training:}
During training, the waveform model encodes acoustic features $X$ into the latent acoustic representation $z_X$ and decodes the waveform from $z_X$. The latent acoustic model predicts $z_X$ given the latent linguistic representation $z_Y$ as input, which is encoded by the linguistic encoder from the linguistic feature $Y$. The alignment model learns to align the latent acoustic and linguistic representations, and the duration model is trained to predict the duration derived from the alignment model. The training is divided into two stages: in the first stage, the waveform model is trained independently; in the second stage, the other models are trained with the parameters of waveform model fixed.

\noindent 
\textbf{Inference:}
During inference, latent linguistic representations $z_Y$ encoded by the linguistic encoder given the linguistic feature $Y$ are upsampled on the basis of the duration predicted by the duration model given $z_Y$ as input. The latent acoustic model predicts latent acoustic representations $z_X$ from the upsampled $z_Y$. Finally, the waveform model predicts the waveform from $z_X$.
\vspace{-3mm}
\subsection{Model components}
\vspace{-2mm}
\noindent 
\textbf{Waveform model:}
The waveform model consists of the acoustic encoder and waveform decoder. The acoustic encoder encodes acoustic features $X$ into the latent acoustic  representation $z_X$, which follows the Gaussian distribution with the approximated mean $\mu_\psi(X)$ and variance $\sigma_\psi^2(X)$. The waveform decoder decodes the waveform given $z_X$. In the framework of DPM, the waveform model can be regarded as the data distribution model $p(x_0|x_1)$ in Eq.~(\ref{eq:diff_0}) considering $x_0$ as the waveform and $x_1$ as $z_X$. 

\noindent 
\textbf{Latent acoustic model:}
The latent acoustic model is the core of the TTS model to represent the conditional distribution $p(z_X|z_Y)$ that predicts the output latent acoustic representation $z_X$ given the input latent linguistic representation $z_Y$. $z_Y$ is obtained from the linguistic encoder by encoding linguistic features $Y$. $z_X$ is obtained from the waveform model.
We model the conditional distribution with diffusion.
We design to diffuse the mean from 0 to $\mu_\psi(X)$ and the variance from 1 to $\sigma_\psi^2(X)$, whereas common diffusion models diffuse the mean from 0 to $x_0$ and the variance from 1 to 0 as $t$ runs from $T$ to 0. We take this approach because the distribution of $z_X$ is known, as approximated by the acoustic encoder. 
To define the approximate posterior, we treat $\mu_\psi(X)$ as the target data $x_0$ to determine the mean parameter in Eq.~(\ref{eq:mean-approx-poasterior}).
To derive the variance of the approximate posterior, we interpolate variance  $\bar{\sigma}^2(X,t)$ from $\sigma_\psi^2(X)$ to 1 as $\bar{\sigma}^2(X,t) = \bar{\beta}_t + (1 - \bar{\beta}_t)\sigma^2_\psi(X)$ on the basis of Eq.~(\ref{eq:variance-approx-poasterior}). 
We assume the prior distribution $p(x_T)$ to be a standard Gaussian distribution. The mean and variance parameters of the model distribution $p_\theta(x_{t-1}|x_t)$ are predicted by the model function $f_\theta(x_t, t, z_Y)$, and they are optimized to be close to the approximate posterior by minimizing the KL divergence in Eq.~(\ref{eq:diff_t}). We choose the mean $\mu_\psi(X)$ and variance $\sigma_\psi^2(X)$ of $z_X$ as the target of the model function. We also attempted noise prediction with the model function in preliminary experiments, but the quality of synthetic speech was worse than the mean parameter prediction, which was consistent with the findings in \cite{DBLP:journals/corr/abs-2207-06389, DBLP:journals/corr/abs-2204-06125}. 

\noindent 
\textbf{Alignment model:}
The alignment model is used to learn alignments between the latent linguistic $z_Y$ and acoustic $z_X$ representations. We introduce a function $g_\phi$ to convert $z_X$ into $z_Y$ to align them. To obtain the function $g_\phi$ that satisfies $z_Y = g_\phi(z_X)$, we optimize the parameter $\phi$ by minimizing the square distance between $z_Y$ and $g_\phi(z_X)$. We treat $z_Y$ and $z_X$ as constants to optimize $g_\phi$ to prevent them from collapsing to meaningless representations. Alignments are derived as the monotonic path with the least total distance within trellis defined by the square distance between $z_Y$ and $g_\phi(z_X)$. 
Duration can be derived by summing the number of acoustic frames per linguistic unit from alignments. 
The duration model is trained using duration labels obtained from the alignment model. 
\vspace{-4mm}
\subsection{Comparison with related methods}
\vspace{-2mm}
VITS \cite{DBLP:conf/icml/KimKS21} and Grad-TTS \cite{DBLP:conf/icml/PopovVGSK21} are representative methods related to ours. VITS consists of a waveform model based on VAE and an acoustic model that predicts the distribution of the latent representation of the waveform model, which is a similar framework to that in our method. The differences between VITS and our proposed method are as follows: (1) VITS uses a flow-based probabilistic model \cite{rezende2015variational} for the latent acoustic model; (2) the invertible function in the flow can act as the alignment model by learning the conversion between acoustic and linguistic representations; (3) the waveform model of VITS is jointly trained with the latent acoustic model. 

Grad-TTS uses DPM similarly to our method, but it has the following differences: (1) it models the mel-spectrogram as the output instead of latent acoustic representations; (2) the target of the model function in diffusion is noise instead of the mean parameter; (3) its prior models the mel-spectrogram instead of a standard Gaussian noise; (4) its alignments are derived from trellis defined by distances between prior's predictions and ground truth mel-spectrograms. 

\vspace{-4mm}
\section{Experimental Evaluations}
\label{sec:experiments}
\vspace{-3mm}
\subsection{Experimental conditions}
\vspace{-3mm}
To construct our proposed TTS systems, we trained an identical waveform model shared across all the proposed systems. We then trained acoustic and alignment models using the trained waveform model. We used the WaveNet \cite{oord2016wavenet} structure for the acoustic encoder, and we used HiFi-GAN \cite{DBLP:conf/nips/KongKB20} for the waveform decoder in the waveform model. We used the WaveNet \cite{oord2016wavenet} structure to implement the latent acoustic and alignment models. We used the convolutional neural network to implement the duration model. We used the self-attention-based linguistic encoder \cite{DBLP:conf/icml/KimKS21}. In the latent acoustic model, we set the number of diffusion steps to be 100.
We sampled diffusion time $t$ uniformly.
We optimized KL divergence in Eq.~(\ref{eq:diff_t}) directly, although it was reported to be difficult \cite{DBLP:conf/nips/HoJA20, DBLP:conf/icml/NicholD21}.

Table \ref{tbl:systems} shows the TTS systems we investigated. We included three proposed systems with different alignment conditions to evaluate our alignment method. The first system used alignments predicted from our method to train the duration and latent acoustic models (\texttt{DFPA}). We used monotonic alignment search \cite{DBLP:conf/nips/KimKKY20} to search for monotonic alignments in the trellis. The second system used ground truth alignments to train the duration and latent acoustic models (\texttt{DFVA}). The third system used ground truth alignments during both training and inference without using the duration model (\texttt{DFVAA}). We used alignments obtained from VITS \cite{DBLP:conf/icml/KimKS21} as ground truth alignments.

We tested two latent acoustic models to evaluate the effect of the probabilistic model for our proposed method. The first variant used DPM by optimizing the latent acoustic model with Eq.~(\ref{eq:diff_t}) (\texttt{DFPA}, \texttt{DFVA}, \texttt{DFVAA}). The second system optimized the latent acoustic model with the mean square error (MSE) to predict the mean parameter of acoustic representation (\texttt{DFVAT}). For this system, the isotropic Gaussian distribution of output latent variables was assumed and we attempted to model the temporal correlation of the latent sequences using the Transformer network \cite{Vaswani2017}; this was an approach similar to FastSpeech \cite{DBLP:journals/corr/abs-2006-04558}. In this system, we used ground truth alignments.

As baseline systems, we used VITS \cite{DBLP:conf/icml/KimKS21} and Grad-TTS \cite{DBLP:conf/icml/PopovVGSK21}.
For Grad-TTS, we set the number of diffusion steps to be 100 and changed its network structure from U-Net \cite{DBLP:conf/miccai/RonnebergerFB15} to WaveNet \cite{oord2016wavenet} for higher memory efficiency.

We constructed phoneme and character variants for all TTS methods. We trained a model of our proposed method using phonemes (\texttt{DFPA}) up to 170k steps, a model using phonemes and ground truth alignments (\texttt{DFVA}) up to 92k steps, and a model using characters (\texttt{DFPAC}) up to 115k steps with a batch size of 100. We trained a VITS model using phonemes (\texttt{VSPA}) up to 275k steps and a VITS model using characters up to 67k (\texttt{VSPAC}), with a batch size of 24. We trained a Grad-TTS model using phonemes (\texttt{GTPA}) up to 98k steps, and a Grad-TTS model using characters (\texttt{GTPAC}) up to 90k steps with batch size of 64. The stopping criteria of the training were based on validation loss.

We used the LJSpeech corpus 
to train the TTS systems. LJSpeech is an English corpus consisting of 13,100 utterances from a female speaker. We split these utterances into 12,500, 100, and 400 utterances to construct the training, validation, and test sets. 

We conducted a listening test to evaluate the naturalness of synthetic speech. The test consisted of 13 systems including five proposed systems and four baseline systems plus natural samples and three analysis-by-synthesis (ABS) systems. The ABS systems were waveform models synthesizing with ground truth acoustic features for the proposed systems (\texttt{ABSD}), VITS (\texttt{ABSV}), and Grad-TTS (\texttt{ABSG}). We recruited 265 Japanese listeners by crowdsourcing and obtained 26,000 evaluations. The listeners were asked to rate samples using the five scale mean opinion score (MOS). We tested the statistical significance using the Mann-Whitney U-test \cite{Mann-WhitneyRankTest}.

\begin{table}[t]
\vspace{-4mm}
\caption{TTS systems}
\vspace{-5mm}
\label{tbl:systems}
\begin{center}
\footnotesize
\begin{tabular}{|lccccc|}\hline
\multirow{2}{*}{System}& \multirow{2}{*}{Method}& Acoustic & \multicolumn{2}{c}{Alignment} & \multirow{2}{*}{Input}\\
~ & ~ & model & Training & Inference & ~\\\hline
\texttt{DFPA} & Prop. & Diffusion & Pred.& Pred. & Phoneme\\
\texttt{DFVA} & Prop. & Diffusion & GT& Pred. & Phoneme\\
\texttt{DFVAA} & Prop. & Diffusion & GT& GT & Phoneme\\
\texttt{DFVAT} & Prop. & MSE & GT& Pred. & Phoneme\\
\texttt{DFPAC} & Prop.  & Diffusion & Pred.& Pred. & Character\\
\texttt{VSPA} & VITS & Flow & Pred.& Pred. & Phoneme\\
\texttt{VSPAC} & VITS & Flow & Pred.& Pred. & Character\\
\texttt{GTPA} & Grad-TTS & Diffusion & Pred.& Pred. & Phoneme\\
\texttt{GTPAC} & Grad-TTS & Diffusion & Pred.& Pred. & Character\\
\hline
\end{tabular}
\end{center}
\vspace{-8mm}
\end{table}

\vspace{-3mm}
\subsection{Experimental Results}
\begin{figure}[!t]
    \begin{center}
    {\includegraphics[width=0.8\columnwidth]{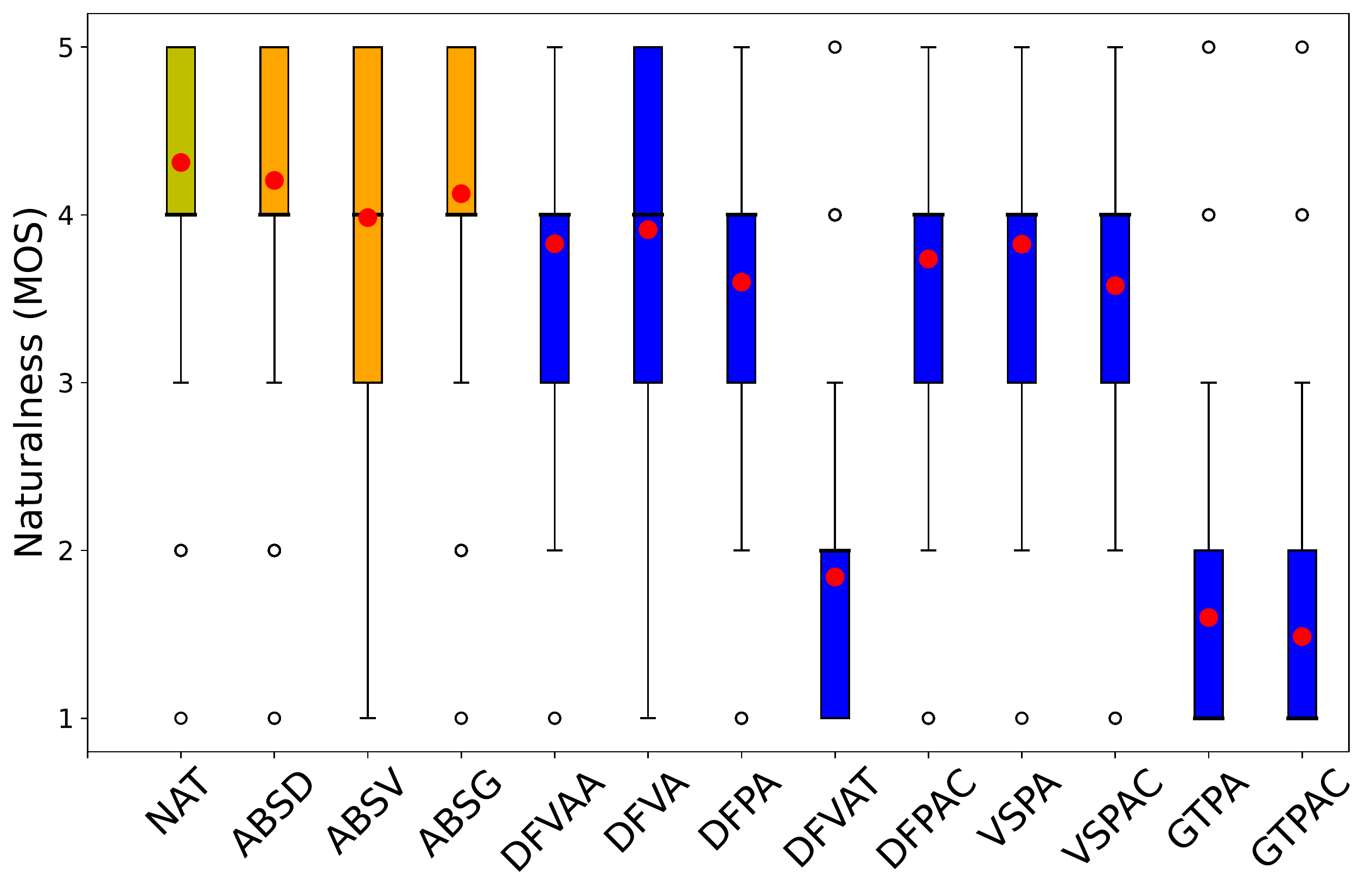}}
    \end{center}
    \vspace{-6mm}
    \caption{Result of listening test on naturalness.}
    \label{fig:mos}
    \vspace{-6mm}
\end{figure}

\vspace{-2mm}
Figure \ref{fig:mos} shows the results of the listening test. The system using the proposed alignment method (\texttt{DFPA}) received a MOS of $3.60\pm0.03$, which was lower than that of the system with ground truth alignments (\texttt{DFVA}) which showed a MOS of $3.91\pm0.03$. This indicated that the accuracy of our alignments method was inferior to that of VITS. The system using ground truth alignment during both training and inference (\texttt{DFVAA}) showed a MOS of $3.83\pm0.03$, which was slightly lower than that of the system using the duration model (\texttt{DFVA}). This indicates that the prediction accuracy of the duration model was sufficient.

As for the probabilistic model, the system replacing DPM with MSE had a low MOS of $1.84\pm0.03$. This indicates that it was difficult to model the latent acoustic distribution with a simple probability distribution assumption even when a powerful network structure such as Transformer was used.

Our method (\texttt{DFPA}) showed a lower MOS than VITS (\texttt{VSPA}) when the proposed alignment method was used: \texttt{DFPA} had a MOS of $3.60\pm0.03$, whereas \texttt{VSPA} had a MOS of $3.83\pm0.03$. In contrast, when ground truth alignments were used, our proposed method (\texttt{DFVA}) achieved a higher score than VITS: \texttt{DFVA} had a higher MOS of $3.91\pm0.03$ than \texttt{VSPA}, which showed a MOS of $3.83\pm0.03$.
We suspected that one of the reasons for the poor performance of our alignment method was the lack of acoustic distribution modeling unlike VITS where its flow-based acoustic model also worked as the alignment model, but further investigation of the alignment method including  VITS is required\footnote{It is suspected that duration derived from VITS are not similar to manually annotated duration:  https://github.com/jaywalnut310/vits/issues/9.}.
Our proposed method benefited from the high naturalness of the waveform model: ABS for our methods (\texttt{ABSD}) had a MOS of $4.21\pm0.03$, whereas that for VITS (\texttt{ABSV}) had a MOS of $3.98\pm0.03$. We inferred that the low naturalness of the VITS waveform model was caused by joint training, as originally reported in their paper \cite{DBLP:conf/icml/KimKS21}, and the independent training of our waveform model gave more consistent quality. VITS showed a MOS closer to that of ABS than our method, which indicated that the acoustic model of VITS performed well or latent features from the VITS waveform model were more predictable.

Grad-TTS showed a significantly low naturalness: \texttt{GTPA} had a MOS of $1.60\pm0.02$. We suspected that the low naturalness was caused by the difficulty in optimizing the model. We found that training a Grad-TTS model long enough to achieve good quality was difficult because its prior model tended to overfit, although the researchers who developed of Grad-TTS stated that long training was essential to achieve good quality \cite{DBLP:conf/icml/PopovVGSK21}\footnote{A Grad-TTS model in the original paper was trained about five times longer than ours, which was 17000k steps with batch size of 16.}. 
Grad-TTS adopts a prior model to predict the mel-spectrogram and a model function to predict noise, which is different from our method using the standard Gaussian prior and model function to predict the mean and variance parameters of the target. Noise was expected to be more difficult to predict than data considering its randomness, and we confirmed it in our preliminary experiment for our method. It is known that noise prediction requires more diffusion steps to generate good samples than data prediction \cite{DBLP:journals/corr/abs-2207-06389}, which implies that noise prediction requires more training steps than data prediction. Our method did not suffer from the slow convergence or overfitting of the prior model.

\begin{figure}[!t]
\vspace{-2mm}
    \begin{center}
    {\includegraphics[width=0.7\columnwidth]{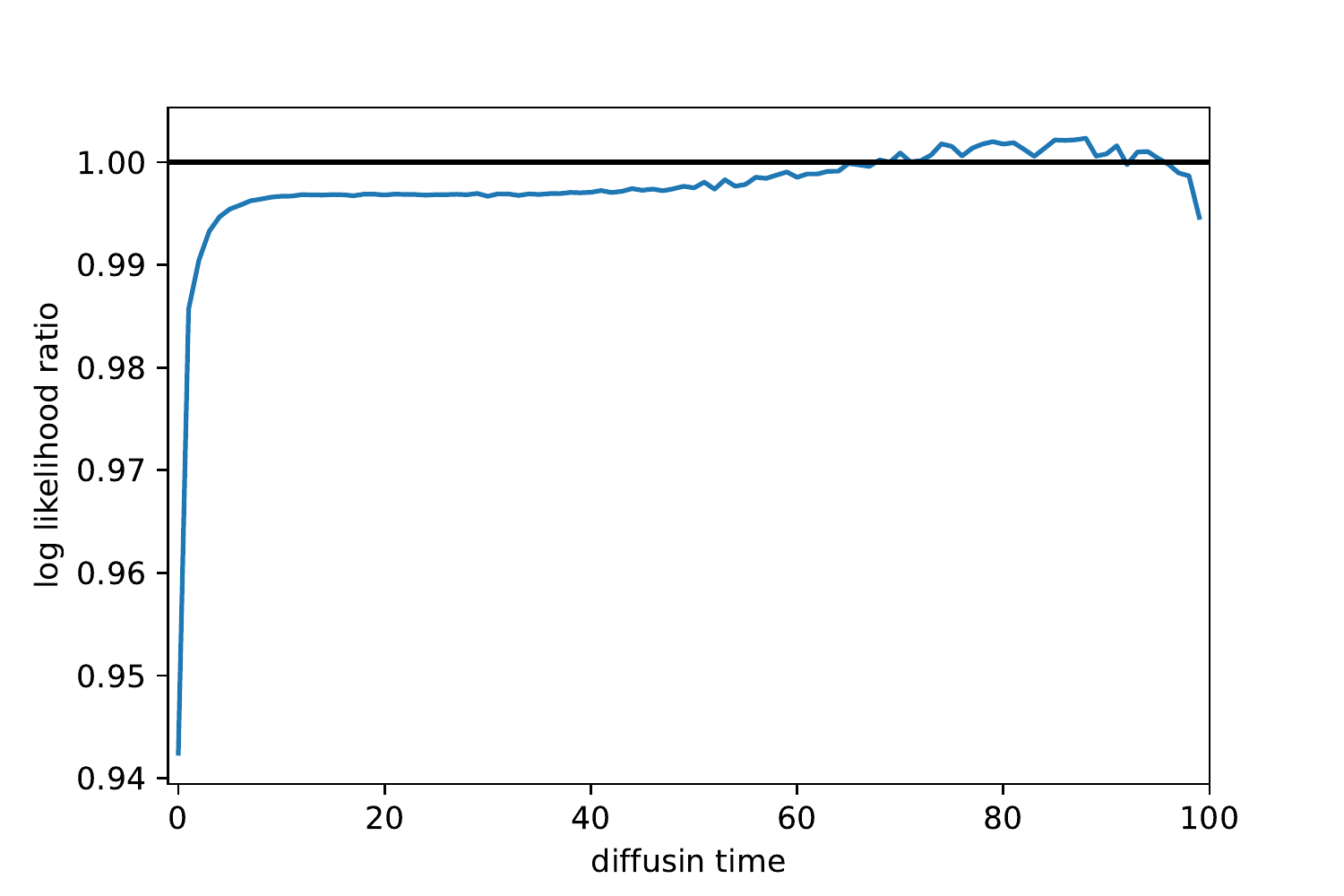}}
    \end{center}
    \vspace{-6mm}
    \caption{Log-likelihood ratio of unconditional to phoneme-conditioned diffusion TTS model.}
    \label{fig:likelihood-ratio}
    \vspace{-6mm}
\end{figure}

As for the differences between phonemes and characters, our proposed method showed different tendencies from the baseline systems. Our method using characters showed improvement of MOS over phonemes, whereas both VITS and Grad-TTS using characters showed degradation. Our method using phonemes (\texttt{DFPA}) had a MOS of $3.60\pm0.03$, whereas our method using characters (\texttt{DFPAC}) had a MOS of $3.74\pm0.03$. VITS using phonemes (\texttt{VSPA}) had a MOS of $3.83\pm0.03$, whereas that using characters (\texttt{VSPAC}) had MOS of $3.58\pm0.03$. Grad-TTS using phonemes (\texttt{GTPA}) had a MOS of $1.60\pm0.02$, whereas that using characters (\texttt{GTPAC}) had a MOS of $1.49\pm0.02$. These results indicated that our proposed methods were robust to linguistic labels with poor orthography. These result were consistent with the relatively high naturalness of synthetic speech obtained by our method, although the accuracy of our alignment method was low. We consider that the diffusion models were robust to the quality of condition labels because the model function in diffusion could use not only conditions but also noisy data $x_t$. To quantitatively analyze the DPM dependence on linguistic labels and noisy data, we measured the log-likelihood ratio of the unconditional to phoneme-conditioned diffusion TTS model in test set as shown in Fig.~\ref{fig:likelihood-ratio}. The ratio $> 1$ indicates that the model depends on linguistic labels over noisy data, whereas the ratio $< 1$ indicates vice versa. We see that the model depended on phonemes over noisy data during large diffusion times, whereas it depended on noisy data over phonemes during small diffusion times. This result was reasonable because during the early inference stage, noisy data consisted of almostly noise and the model had to depend on phonemes, whereas during the late inference stage, noisy data was almost close to the target and the model did not have to use phonemes. Our method (\texttt{DFPAC}) achieved a higher MOS than VITS (\texttt{VSPAC}) under character inputs. This result also supported the stable performance of our proposed method.

\vspace{-5mm}
\section{Conclusion}
\label{sec:conclusion}
\vspace{-3mm}
We proposed a text-to-speech (TTS) method based on latent variable conversion. Our method encodes latent acoustic and linguistic representations using variational autoencoder, and the two latent representations were converted using the diffusion-based acoustic model and the aligned with alignment model. We modeled the approximated mean and variance parameters of the latent representation with diffusion instead of a using fixed variance to incorporate diffusion to latent variable conversion. We evaluated our method along with TTS based on latent variable conversion with flow (VITS) and diffusion-based TTS (Grad-TTS) as baselines. Our method showed the degradation of naturalness due to low alignment accuracy, but outperformed baselines in both phoneme and character conditions when ground truth alignments were given. Our proposed method was robust to alignment errors and linguistic labels with poor orthography because of diffusion, and we performed quantitative analysis to explain it.

Our future work includes the improvement of the alignment method. It is also interesting to investigate self-supervised-based latent representations in our TTS framework.

\vfill\pagebreak
\bibliographystyle{IEEEbib}
\bibliography{reference.bib}

\begin{thebibliography}{10}

\bibitem{DBLP:conf/icassp/ShenPWSJYCZWRSA18_short}
J.~Shen, R.~Pang, R.~J. Weiss, M.~Schuster, N.~Jaitly, Z.~Yang, Z.~Chen,
  Y.~Zhang, Y.~Wang, RJS. Ryan, R.~A. Saurous, Y.~Agiomyrgiannakis, and Y.~Wu,
\newblock ``Natural {TTS} synthesis by conditioning wavenet on {MEL}
  spectrogram predictions,''
\newblock in {\em {ICASSP}}. 2018, pp. 4779--4783, {IEEE}.

\bibitem{DBLP:journals/corr/abs-1809-08895}
Naihan Li, Shujie Liu, Yanqing Liu, Sheng Zhao, Ming Liu, and Ming Zhou,
\newblock ``Close to human quality {TTS} with transformer,''
\newblock {\em CoRR}, vol. abs/1809.08895, 2018.

\bibitem{DBLP:conf/icml/KimKS21}
Jaehyeon Kim, Jungil Kong, and Juhee Son,
\newblock ``Conditional variational autoencoder with adversarial learning for
  end-to-end text-to-speech,''
\newblock in {\em {ICML}}. 2021, vol. 139, pp. 5530--5540, {PMLR}.

\bibitem{oord2016wavenet}
Aaron van~den Oord, Sander Dieleman, Heiga Zen, Karen Simonyan, Oriol Vinyals,
  Alex Graves, Nal Kalchbrenner, Andrew Senior, and Koray Kavukcuoglu,
\newblock ``{Wavenet: A generative model for raw audio},''
\newblock {\em arXiv preprint arXiv:1609.03499}, 2016.

\bibitem{rezende2015variational}
Danilo Rezende and Shakir Mohamed,
\newblock ``Variational inference with normalizing flows,''
\newblock in {\em Proc. ICML}, 2015, pp. 1530--1538.

\bibitem{goodfellow2014generative}
Ian Goodfellow, Jean Pouget-Abadie, Mehdi Mirza, Bing Xu, David Warde-Farley,
  Sherjil Ozair, Aaron Courville, and Yoshua Bengio,
\newblock ``Generative adversarial nets,''
\newblock in {\em Proc. NIPS}, 2014, pp. 2672--2680.

\bibitem{DBLP:conf/nips/HoJA20}
Jonathan Ho, Ajay Jain, and Pieter Abbeel,
\newblock ``Denoising diffusion probabilistic models,''
\newblock in {\em NeurIPS}, 2020.

\bibitem{DBLP:conf/iclr/ChenZZWNC21}
Nanxin Chen, Yu~Zhang, Heiga Zen, Ron~J. Weiss, Mohammad Norouzi, and William
  Chan,
\newblock ``{WaveGrad}: Estimating gradients for waveform generation,''
\newblock in {\em {ICLR}}. 2021, OpenReview.net.

\bibitem{jeong21_interspeech}
Myeonghun Jeong, Hyeongju Kim, Sung~Jun Cheon, Byoung~Jin Choi, and Nam~Soo
  Kim,
\newblock ``{Diff-TTS: A Denoising Diffusion Model for Text-to-Speech},''
\newblock in {\em Proc. Interspeech 2021}, 2021, pp. 3605--3609.

\bibitem{chen21p_interspeech}
Nanxin Chen, Yu~Zhang, Heiga Zen, Ron~J. Weiss, Mohammad Norouzi, Najim Dehak,
  and William Chan,
\newblock ``{WaveGrad 2: Iterative Refinement for Text-to-Speech Synthesis},''
\newblock in {\em Proc. Interspeech 2021}, 2021, pp. 3765--3769.

\bibitem{DBLP:conf/iclr/LeeKS0LMQ0YL22}
Sang{-}gil Lee, Heeseung Kim, Chaehun Shin, Xu~Tan, Chang Liu, Qi~Meng, Tao
  Qin, Wei Chen, Sungroh Yoon, and Tie{-}Yan Liu,
\newblock ``{PriorGrad}: Improving conditional denoising diffusion models with
  data-dependent adaptive prior,''
\newblock in {\em {ICLR}}. 2022, OpenReview.net.

\bibitem{DBLP:conf/icml/KimKY22}
Heeseung Kim, Sungwon Kim, and Sungroh Yoon,
\newblock ``{Guided-TTS}: {A} diffusion model for text-to-speech via classifier
  guidance,''
\newblock in {\em {ICML}}. 2022, vol. 162, pp. 11119--11133, {PMLR}.

\bibitem{levkovitch22_interspeech}
Alon Levkovitch, Eliya Nachmani, and Lior Wolf,
\newblock ``{Zero-Shot Voice Conditioning for Denoising Diffusion TTS
  Models},''
\newblock in {\em Proc. Interspeech 2022}, 2022, pp. 2983--2987.

\bibitem{DBLP:journals/corr/KingmaW13}
Diederik~P. Kingma and Max Welling,
\newblock ``Auto-encoding variational bayes,''
\newblock in {\em {ICLR}}, 2014.

\bibitem{DBLP:conf/interspeech/KlimkovRRD19}
Viacheslav Klimkov, Srikanth Ronanki, Jonas Rohnke, and Thomas Drugman,
\newblock ``Fine-grained robust prosody transfer for single-speaker neural
  text-to-speech,''
\newblock in {\em {INTERSPEECH}}. 2019, pp. 4440--4444, {ISCA}.

\bibitem{DBLP:conf/iclr/HsuZWZWWCJCSNP19}
Wei{-}Ning Hsu, Yu~Zhang, Ron~J. Weiss, Heiga Zen, Yonghui Wu, Yuxuan Wang,
  Yuan Cao, Ye~Jia, Zhifeng Chen, Jonathan Shen, Patrick Nguyen, and Ruoming
  Pang,
\newblock ``Hierarchical generative modeling for controllable speech
  synthesis,''
\newblock in {\em {ICLR} (Poster)}. 2019, OpenReview.net.

\bibitem{DBLP:conf/nips/OordVK17}
A{\"{a}}ron van~den Oord, Oriol Vinyals, and Koray Kavukcuoglu,
\newblock ``Neural discrete representation learning,''
\newblock in {\em {NIPS}}, 2017, pp. 6306--6315.

\bibitem{DBLP:conf/iclr/HabibMSBSSKB20}
Raza Habib, Soroosh Mariooryad, Matt Shannon, Eric Battenberg, R.~J.
  Skerry{-}Ryan, Daisy Stanton, David Kao, and Tom Bagby,
\newblock ``Semi-supervised generative modeling for controllable speech
  synthesis,''
\newblock in {\em {ICLR}}. 2020, OpenReview.net.

\bibitem{jia21_interspeech}
Ye~Jia, Heiga Zen, Jonathan Shen, Yu~Zhang, and Yonghui Wu,
\newblock ``{PnG BERT: Augmented BERT on Phonemes and Graphemes for Neural
  TTS},''
\newblock in {\em Proc. Interspeech 2021}, 2021, pp. 151--155.

\bibitem{DBLP:conf/interspeech/SiuzdakDRJ22}
Hubert Siuzdak, Piotr Dura, Pol van Rijn, and Nori Jacoby,
\newblock ``{WavThruVec}: Latent speech representation as intermediate features
  for neural speech synthesis,''
\newblock in {\em {INTERSPEECH}}. 2022, pp. 833--837, {ISCA}.

\bibitem{9829304}
Yusuke Yasuda and Tomoki Toda,
\newblock ``Investigation of japanese {PnG BERT} language model in
  text-to-speech synthesis for pitch accent language,''
\newblock {\em IEEE Journal of Selected Topics in Signal Processing}, vol. 16,
  no. 6, pp. 1319--1328, 2022.

\bibitem{DBLP:conf/icml/NicholD21}
Alexander~Quinn Nichol and Prafulla Dhariwal,
\newblock ``Improved denoising diffusion probabilistic models,''
\newblock in {\em {ICML}}. 2021, vol. 139, pp. 8162--8171, {PMLR}.

\bibitem{DBLP:journals/corr/abs-2207-06389}
Rongjie Huang, Zhou Zhao, Huadai Liu, Jinglin Liu, Chenye Cui, and Yi~Ren,
\newblock ``Prodiff: Progressive fast diffusion model for high-quality
  text-to-speech,''
\newblock {\em CoRR}, vol. abs/2207.06389, 2022.

\bibitem{DBLP:conf/nips/SongE19}
Yang Song and Stefano Ermon,
\newblock ``Generative modeling by estimating gradients of the data
  distribution,''
\newblock in {\em NeurIPS}, 2019, pp. 11895--11907.

\bibitem{DBLP:journals/corr/abs-2204-06125}
Aditya Ramesh, Prafulla Dhariwal, Alex Nichol, Casey Chu, and Mark Chen,
\newblock ``Hierarchical text-conditional image generation with {CLIP}
  latents,''
\newblock {\em CoRR}, vol. abs/2204.06125, 2022.

\bibitem{DBLP:conf/icml/PopovVGSK21}
Vadim Popov, Ivan Vovk, Vladimir Gogoryan, Tasnima Sadekova, and Mikhail~A.
  Kudinov,
\newblock ``{Grad-TTS}: {A} diffusion probabilistic model for text-to-speech,''
\newblock in {\em {ICML}}. 2021, vol. 139, pp. 8599--8608, {PMLR}.

\bibitem{DBLP:conf/nips/KongKB20}
Jungil Kong, Jaehyeon Kim, and Jaekyoung Bae,
\newblock ``{HiFi-GAN}: Generative adversarial networks for efficient and high
  fidelity speech synthesis,''
\newblock in {\em NeurIPS}, 2020.

\bibitem{DBLP:conf/nips/KimKKY20}
Jaehyeon Kim, Sungwon Kim, Jungil Kong, and Sungroh Yoon,
\newblock ``{Glow-TTS}: {A} generative flow for text-to-speech via monotonic
  alignment search,''
\newblock in {\em NeurIPS}, 2020.

\bibitem{Vaswani2017}
Ashish Vaswani, Noam Shazeer, Niki Parmar, Jakob Uszkoreit, Llion Jones,
  Aidan~N. Gomez, Lukasz Kaiser, and Illia Polosukhin,
\newblock ``Attention is all you need,''
\newblock in {\em Proc. NIPS}, 2017, pp. 6000--6010.

\bibitem{DBLP:journals/corr/abs-2006-04558}
Yi~Ren, Chenxu Hu, Xu~Tan, Tao Qin, Sheng Zhao, Zhou Zhao, and Tie{-}Yan Liu,
\newblock ``Fastspeech 2: Fast and high-quality end-to-end text to speech,''
\newblock {\em CoRR}, vol. abs/2006.04558, 2020.

\bibitem{DBLP:conf/miccai/RonnebergerFB15}
Olaf Ronneberger, Philipp Fischer, and Thomas Brox,
\newblock ``U-net: Convolutional networks for biomedical image segmentation,''
\newblock in {\em {MICCAI} {(3)}}. 2015, vol. 9351 of {\em Lecture Notes in
  Computer Science}, pp. 234--241, Springer.

\bibitem{Mann-WhitneyRankTest}
H.~B. Mann and D.~R. Whitney,
\newblock ``On a test of whether one of two random variables is stochastically
  larger than the other,''
\newblock {\em The Annals of Mathematical Statistics}, vol. 18, no. 1, pp.
  50--60, 1947.

\end{thebibliography}

\end{document}